\magnification=\magstep1
\hsize=125mm
\overfullrule=0pt
\font\title=cmbx10 scaled\magstep2

                                                  JYFL preprint 17/1997

\bigskip

\bigskip

\bigskip

\bigskip

\bigskip

\bigskip

\bigskip

\bigskip

\bigskip

\centerline{\title Variation of Area Variables in Regge Calculus}

\bigskip

\bigskip

\centerline{Jarmo M\"{a}kel\"{a}\footnote{$^1$}{e-mail: makela@jyfl.jyu.fi}}

\bigskip

\centerline{\it Department of Physics, University of Jyv\"{a}skyl\"{a}, P. O.
Box 35, FIN-40351}

\centerline{\it Jyv\"{a}skyl\"{a}, Finland}

\bigskip

\bigskip

\centerline{\bf Abstract}

\bigskip

     We consider the possibility to use the areas of two-simplexes, instead of
lengths of edges, as the dynamical variables of Regge calculus. We show that if
the action of Regge calculus is varied with respect to the areas of
two-simplexes, and appropriate constraints are imposed between the variations,
the Einstein-Regge equations are recovered.

\bigskip


\vfill\eject

       Regge calculus is an approach to general relativity where spacetime is
modelled by a piecewise flat, or simplicial, manifold. More precisely, a
four-dimensional geometrical simplicial complex is used as a model of
four-dimensional spacetime, and the geometrical properties of that manifold
are described in terms of the edge lengths of the complex. In other words,
Regge calculus involves a specific discrete theory of space and time where edge
lengths are the fundamental dynamical variables.[1-3] 

      Although edge lengths are the fundamental variables in Regge calculus, it
is very interesting to investigate the possibility to use {\it areas of
two-simplexes}, instead of lengths of edges, as the dynamical variables. This
sort of an approach is motivated, among other things, by the success met by the 
so called loop quantum gravity.[4] In loop quantum gravity, the quantum states of
spacetime are associated with certain kind of loops lying on a spacelike
hypersurface of spacetime. Hence, we are suggested a view that one should use
quantities associated with loop-like objects as the fundamental
variables of quantum gravity. In Regge calculus, the loops are two-simplexes,
or triangles, and the natural loop variables are their areas. 

       The use of areas as independent dynamical variables in Regge calculus
has already received an attention of some authors.[5-7] However, it appears that
the theory emerging as an outcome of such an approach is different from the
original Regge calculus approach . This result is due to the fact that the number of
triangles in a four-complex is, in general, different from the number of edges.
For instance, the number of edges on a four-simplex is
$$
\left(\matrix{5\cr
              2\cr}\right) = 10,\eqno(1)
$$
which is the same as the number of triangles:
$$
\left(\matrix{5\cr
              3\cr}\right) = 10.\eqno(2)
$$
However, if two four-simplexes are clued together along their common
three-simplex, the number of remaining edges is
$$
2\times\left(\matrix{5\cr
                     2\cr}\right) - \left(\matrix{4\cr
                                                  2\cr}\right) = 14,\eqno(3)
$$
but the number of remaining triangles is
$$
2\times\left(\matrix{5\cr
                     3\cr}\right) - \left(\matrix{4\cr
                                                  3\cr}\right) = 16,\eqno(4)
$$
In other words, the number of remaining triangles is {\it greater} than the
number of remaining edges. In general, the number of triangles in any closed
simplicial four-manifold is at least $4\slash 3$ times the number of edges.[7]
This means that if one
wants to recover the original Regge calculus approach by using areas of 
triangles
as the dynamical variables, then all of these variables are not independent of
each other but there are certain {\it constraints} between them. 

        The problem is then to find these constraints. A clue to the solution
of this problem could perhaps be found by trying to find an expression to the 
edge lengths of a four-simplex in terms of the areas of its triangles. From the
outset, this might seem possible since, as we saw in Eqs.(1) and (2), the
number of edges of a four-simplex is the same as the number of its triangles.
Unfortunately, the relationship between the areas and edge lengths is pretty
complicated. More precisely, the area of a triangle with vertices $v_\alpha$,
$v_\beta$ and $v_\gamma$ is, in terms of its squared edge lengths $s_{\alpha\beta}$,
$s_{\alpha\gamma}$ and $s_{\beta\gamma}$,
$$
A_{\alpha\beta\gamma} = {1\over 4}\sqrt{4s_{\alpha\beta}s_{\beta\gamma} -
(s_{\alpha\beta} + s_{\beta\gamma} - s_{\alpha\gamma})^2}.\eqno(5)
$$
Because of the complexity of this expression, one expects that it is not
possible to express, in a closed form, the edge lenghts of a four-simplex 
in terms of the areas of its two-simplexes. Hence, one expects that
it is not possible to express, in a closed form, the constraints between areas
either.

      The situation is not, however, quite as hopeless as it might seem.
Instead of trying to find the constraints between areas themselves, we should
perhaps try to find the constraints between their {\it variations}. In other
words, we pose ourselves a new problem: In which way should we vary the area
variables in the action of Regge calculus if we want to recover the equations
obtainable by varying the action with respect to the edge lengths?

      Before going into this question, we introduce a new type of notation. One
of the basic ideas of our notation is to identify every simplicial object by 
means of the corresponding vertices. For instance, we associate with every
vertex pair $(v_\mu,v_\nu)$ of our complex $K$ a quantity $s_{\mu\nu}$ such 
that $s_{\mu\nu}$ is the squared length of the one-simplex, or edge, $[v_\mu
v_\nu]$ joining the vertices $v_\mu$ and $v_\nu$ if the one-simplex $[v_\mu
v_\nu]\in K$, and we set $s_{\mu\nu}:=0$ if $[v_\mu v_\nu]\not\in K$. It is
clear that the quantities $s_{\mu\nu}$ are real and non-negative, and they have a
property
$$
s_{\mu\nu} = s_{\nu\mu}\eqno(6)
$$
for every $v_\mu, v_\nu\in K$ (For the sake of simplicity, we assume
Euclidean, rather than Lorentzian spacetime). Moreover, we associate with every vertex triplet
$(v_\alpha,v_\beta,v_\gamma)$ of $K$ a quantity $A_{\alpha\beta\gamma}$ such
that $A_{\alpha\beta\gamma}$ is the area of the two-simplex $[v_\alpha v_\beta
v_\gamma]$ if $[v_\alpha v_\beta v_\gamma]\in K$, and we set
$A_{\alpha\beta\gamma}:=0$, if $[v_\alpha v_\beta v_\gamma]\not\in K$. Again,
$A_{\alpha\beta\gamma}$ is real and non-negative, and it is totally symmetric
with respect to its indices. 

      Another element of our notation is the use of Einstein's summation
convention such that repeated indices up and down are summed over. Unless
otherwise stated, the sum is taken over {\it all} vertices of the complex $K$.
With these notations and conventions the action of Regge calculus --which in the
usual notation is written as ($G=c=1$):
$$
S = -{1\over{8\pi}}\sum_i A_i\phi_i,\eqno(7)
$$
where the sum is taken over two-simplexes of the complex ($A_i$'s are their
areas and $\phi_i$'s the corresponding deficit angles)-- now takes the form:
$$
S = -{1\over{8\pi}}
{1\over{3!}}A_{\alpha\beta\gamma}\phi^{\alpha\beta\gamma},\eqno(8)
$$ 
where $\phi^{\alpha\beta\gamma}$ is the deficit angle corresponding to the
two-simplex $[v_\alpha v_\beta v_\gamma]$. We set
$\phi^{\alpha\beta\gamma}:=0$ if $[v_\alpha v_\beta v_\gamma]\not\in K$.
$\phi^{\alpha\beta\gamma}$ is totally symmetric with respect to its indices,
and the reason for an appearance of the factor ${1\over{3!}}$ is the total
symmetry of the quantities $A_{\alpha\beta\gamma}$ and
$\phi^{\alpha\beta\gamma}$. It should be noted that although the sum in Eq.(8)
is taken over all vertices of the complex $K$, the terms associated with the
vertex triplets $(v_\alpha, v_\beta, v_\gamma)$ yield non-zero contributions to
the sum only if $[v_\alpha v_\beta v_\gamma]\in K$.

            When addressing the question of constraints between the variations
of areas of two-simplexes, the first step is to investigate the effects of
variations of squared edge lengths on the areas of two-simplexes. In general, the
variations of the areas $A_{\alpha\beta\gamma}$ can be written in terms of the
variations of squared edge lengths as:
$$
\delta A_{\alpha\beta\gamma} = M_{\alpha\beta\gamma}^{\mu\nu} \delta
s_{\mu\nu},\eqno(9)
$$
where
$$
M_{\alpha\beta\gamma}^{\mu\nu}:= {1\over 2} {{\partial
A_{\alpha\beta\gamma}}\over{\partial s_{\mu\nu}}},\eqno(10)
$$
for all vertices $v_\mu, v_\nu, v_\alpha, v_\beta, v_\gamma\in K$ (the 
factor ${1\over 2}$ is due to the symmetry of $s_{\mu\nu}$).
It follows from Eq.(5) that
$$
M_{\alpha\beta\gamma}^{\mu\nu} = {1\over 2} \Delta_{\beta\gamma}^\alpha
(\delta_\beta^\mu \delta_\gamma^\nu + \delta_\gamma^\mu \delta_\beta^\nu) +
{1\over 2} \Delta_{\alpha\gamma}^\beta
(\delta_\alpha^\mu \delta_\gamma^\nu + \delta_\gamma^\mu \delta_\alpha^\nu) +
 {1\over 2}\Delta_{\alpha\beta}^\gamma
(\delta_\alpha^\mu \delta_\beta^\nu + \delta_\beta^\mu
\delta_\alpha^\nu),\eqno(11)
$$
where we have defined
$$
\Delta_{\alpha\beta}^\gamma :=
{1\over{16A_{\alpha\beta\gamma}}}(s_{\alpha\gamma} + s_{\beta\gamma} -
s_{\alpha\beta})\eqno(12)
$$
for every non-zero $A_{\alpha\beta\gamma}$. If $A_{\alpha\beta\gamma}=0$, we
set $\Delta_{\alpha\beta}^\gamma:=0$. 

        Is it possible to invert Eq.(9) and to write the variations of squared
edge lengths of the complex in terms of the variations of areas of two-simplexes?
In general, this is not possible since the number of two-simplexes usually is
greater that the number of edges. Under certain conditions, however, it is
possible to express the variations of squared edge lengths of a given {\it
four-simplex} in terms of the variations of the areas of its two-simplexes. To
this order, we pick up a four-simplex $\sigma$ of the complex $K$, and
we associate with the four-simplex $\sigma$ the quantities
$M^{\mu\nu}_{\alpha\beta\gamma}(\sigma)$ such that
$$
M_{\alpha\beta\gamma}^{\mu\nu}(\sigma) 
:=M_{\alpha\beta\gamma}^{\mu\nu},\eqno(13.a)
$$
if $[v_\mu v_\nu v_\alpha v_\beta v_\gamma]=\sigma$, and
$$
M_{\alpha\beta\gamma}^{\mu\nu}(\sigma) :=0,\eqno(13.b)
$$
if $[v_\mu v_\nu v_\alpha v_\beta v_\gamma]\neq\sigma$.
By means of the quantities $M_{\alpha\beta\gamma}^{\mu\nu}(\sigma)$, we can
write the variations of areas $A_{\alpha\beta\gamma}(\sigma)$ of the
two-simplexes of the four-simplex $\sigma$ in terms of the variations 
of its squared edge lengths $s_{\mu\nu}(\sigma)$:
$$
\delta A_{\alpha\beta\gamma}(\sigma) = 
M_{\alpha\beta\gamma}^{\mu\nu}(\sigma) \delta
s_{\mu\nu}(\sigma),\eqno(14)
$$
For a four-simplex $[v_0 v_1 v_2 v_3 v_4]$, we can
write Eq.(14) in a matrix form:
$$
\left(\matrix{\delta A_{012}\cr
                \delta A_{013}\cr
                \delta A_{014}\cr
                \delta A_{023}\cr
                \delta A_{024}\cr
                \delta A_{034}\cr
                \delta A_{123}\cr
                \delta A_{124}\cr
                \delta A_{134}\cr
                \delta A_{234}\cr}\right) = \left(\matrix{
\Delta_{01}^2&\Delta_{02}^1&0&0&\Delta_{12}^0&0&0&0&0&0\cr
\Delta_{01}^3&0&\Delta_{03}^1&0&0&\Delta_{13}^0&0&0&0&0\cr
\Delta_{01}^4&0&0&\Delta_{04}^1&0&0&\Delta_{14}^0&0&0&0\cr
0&\Delta_{02}^3&\Delta_{03}^2&0&0&0&0&\Delta_{23}^0&0&0\cr
0&\Delta_{02}^4&0&\Delta_{04}^2&0&0&0&0&\Delta_{24}^0&0\cr
0&0&\Delta_{03}^4&\Delta_{04}^3&0&0&0&0&0&\Delta_{34}^0\cr
0&0&0&0&\Delta_{12}^3&\Delta_{13}^2&0&\Delta_{23}^1&0&0\cr
0&0&0&0&\Delta_{12}^4&0&\Delta_{14}^2&0&\Delta_{24}^1&0\cr
0&0&0&0&0&\Delta_{13}^4&\Delta_{14}^3&0&0&\Delta_{34}^1\cr
0&0&0&0&0&0&0&\Delta_{23}^4&\Delta_{24}^3&\Delta_{34}^2\cr}\right)
\left(\matrix{\delta s_{01}\cr
              \delta s_{02}\cr
              \delta s_{03}\cr
              \delta s_{04}\cr
              \delta s_{12}\cr
              \delta s_{13}\cr
              \delta s_{14}\cr
              \delta s_{23}\cr
              \delta s_{24}\cr
              \delta s_{34}\cr}\right).\eqno(15)
$$
When writing this equation, we have identified the rows with the triplets
$(\alpha,\beta,\gamma)$, and the columns with the pairs $(\mu,\nu)$. In other
words, the quantities $M_{\alpha\beta\gamma}^{\mu\nu}(\sigma)$ can be understood as
the elements of a certain matrix {\bf M}$(\sigma)$ corresponding to the simplex
$\sigma$.
Since every four-simplex has ten edges
and ten triangles, {\bf M}$(\sigma)$ is a ten by ten matrix.
          
        In what follows, we shall assume that 
$$
det({\hbox{\bf M}}(\sigma))\neq 0\eqno(16)
$$
for every four-simplex $\sigma$ of the complex $K$. For every four-simplex
$\sigma\in K$ we define the quantities $N^{\alpha\beta\gamma}_{\mu\nu}(\sigma)$
such that 
$$
N^{\alpha\beta\gamma}_{\mu\nu}(\sigma) := {{1\over{3!det({\hbox{\bf
M}}(\sigma))}}}C^{\alpha\beta\gamma}_{\mu\nu}(\sigma),\eqno(17.a)
$$
if $\sigma=[v_\mu v_\nu v_\alpha v_\beta v_\gamma]$, and
$$
N^{\alpha\beta\gamma}_{\mu\nu}(\sigma) :=0,\eqno(17.b)
$$
if $\sigma\neq[v_\mu v_\nu v_\alpha v_\beta v_\gamma]$.
In these equations, $C^{\alpha\beta\gamma}_{\mu\nu}(\sigma)$ is the cofactor
corresponding to the row $(\alpha,\beta,\gamma)$ and the column $(\mu,\nu)$ of the
matrix ${\hbox{\bf M}}(\sigma)$. Because of Cramer's rule, the quantities
$N^{\alpha\beta\gamma}_{\mu\nu}(\sigma)$ have a property:
$$
N^{\alpha\beta\gamma}_{\mu\nu}(\sigma)M_{\alpha\beta\gamma}^{\rho\omega}
={1\over 2}(\delta^\rho_\mu \delta^\omega_\nu 
+ \delta^\omega_\mu \delta^\rho_\nu)\eqno(18)
$$
for every four-simplex $\sigma\in K$. Hence, we can write the variations of
squared edge lengths of the four-simplex $\sigma$ in terms of the variations of
the areas of the two-simplexes of the complex:
$$
\delta s_{\mu\nu}(\sigma) = N^{\alpha\beta\gamma}_{\mu\nu}(\sigma)\delta
A_{\alpha\beta\gamma}.\eqno(19)
$$
It should be noted that although the sum is here taken over all vertices of the
complex, only the variations of the areas of the two-simplexes of the
four-simplex $\sigma$ contribute non-vanishing terms. The non-zero components
of $N^{\alpha\beta\gamma}_{\mu\nu}(\sigma)$ can, again, be understood as
elements of a certain ten by ten matrix ${\hbox{\bf N}}(\sigma)$, which is the
inverse of ${\hbox{\bf M}}(\sigma)$. For instance, if the edge lengths of a
four-simplex $[v_0 v_1v_2v_3v_4]$ are equal, Eq.(19) can be written in a matrix
form:\footnote{$^1$}{I am grateful to Pasi Repo for calculating this matrix
form.}
$$
\left(\matrix{\delta s_{01}\cr
              \delta s_{02}\cr
              \delta s_{03}\cr
              \delta s_{04}\cr
              \delta s_{12}\cr
              \delta s_{13}\cr
              \delta s_{14}\cr
              \delta s_{23}\cr
              \delta s_{24}\cr
              \delta s_{34}\cr}\right)
= {{2\sqrt{3}}\over 3}\left(\matrix{2&2&2&-1&-1&-1&-1&-1&-1&2\cr
                                    2&-1&-1&2&2&-1&-1&-1&2&-1\cr
                                    -1&2&-1&2&-1&2&-1&-1&2&-1\cr
                                    -1&-1&2&-1&2&2&2&-1&-1&-1\cr
                                    2&-1&-1&-1&-1&2&2&2&-1&-1\cr
                                    -1&2&-1&-1&2&-1&2&-1&2&-1\cr
                                    -1&-1&2&2&-1&-1&-1&2&2&-1\cr
                                    -1&-1&2&2&-1&-1&2&-1&-1&2\cr
                                    -1&2&-1&-1&2&-1&-1&2&-1&2\cr
                                    2&-1&-1&-1&-1&2&-1&-1&2&2\cr}\right)
\left(\matrix{\delta A_{012}\cr
                \delta A_{013}\cr
                \delta A_{014}\cr
                \delta A_{023}\cr
                \delta A_{024}\cr
                \delta A_{034}\cr
                \delta A_{123}\cr
                \delta A_{124}\cr
                \delta A_{134}\cr
                \delta A_{234}\cr}\right).\eqno(20)
$$

     Consider now an edge joining the vertices $v_\mu$ and $v_\nu$ of the complex
such that this edge is shared by {\it two} four-simplexes, which we shall
denote by $\sigma$ and $\tau$. If the edge in question is considered as one of 
the edges of the four-simplex $\sigma$, the variation $\delta 
s_{\mu\nu}(\sigma)$ of its squared
length is given by Eq.(19). On the other hand, if it is considered as one of
the edges of the four-simplex $\tau$, the variation of its squared length is
$$
\delta s_{\mu\nu}(\tau) = N^{\alpha\beta\gamma}_{\mu\nu}(\tau)\delta
A_{\alpha\beta\gamma}.\eqno(21)
$$
However, we have varied the squared length of the one and the same edge; hence
we require that these two variations must be equal:
$$
\delta s_{\mu\nu}(\sigma) = \delta s_{\mu\nu}(\tau),\eqno(22)
$$
which implies that            
$$
(N^{\alpha\beta\gamma}_{\mu\nu}(\sigma) -  
N^{\alpha\beta\gamma}_{\mu\nu}(\tau))\delta
A_{\alpha\beta\gamma} = 0.\eqno(23)
$$
If we define $C^{\alpha\beta\gamma}_{\mu\nu}(\sigma)$ to be zero for every
$\sigma\neq[v_\mu v_\nu v_\alpha v_\beta v_\gamma]$, we get:
$$            
[det({\hbox{\bf M}}(\tau))C^{\alpha\beta\gamma}_{\mu\nu}(\sigma)-
det({\hbox{\bf M}}(\sigma))C^{\alpha\beta\gamma}_{\mu\nu}(\tau)]\delta
A_{\alpha\beta\gamma} = 0.\eqno(24)
$$            
This is our main result. When we have written Eq.(24) for every edge, and every
pair of four-simplexes sharing that edge of the complex, we have written
the constraints between the variations of the areas of the two-simplexes of
the complex. For instance, if we have two four-simplexes  $[v_0v_1v_2v_3v_4]$, 
and $[v_0v_1v_5v_6v_7]$, such that, initially, all edge lengths of both 
simplexes are 
equal, they share an edge joining the vertices $v_0$ and $v_1$, and Eq.(24)
implies the following constraint between the variations of areas:
$$
\eqalign{&2\delta A_{015} + 2\delta A_{016} + 2\delta A_{017} -\delta
A_{056} - \delta A_{057}- \delta A_{067} - \delta A_{156}\cr 
         &- \delta A_{157} - \delta A_{167} +
2\delta A_{567}
         -2\delta A_{012} - 2\delta A_{013} - 2\delta A_{014} + \delta
A_{023}\cr
         &+ \delta A_{024} + \delta A_{034} + \delta A_{123} 
+ \delta A_{124} + \delta
A_{134} - 2\delta A_{234} 
= 0.\cr}\eqno(25)
$$

     Do the constraints (24) imply the Einstein-Regge equations, the dynamical
equations of Regge calculus? To see that this is the case, consider the action
(8) of Regge calculus.  When varying the action (8) with respect to the areas
$A_{\alpha\beta\gamma}$, which are now considered as the dynamical variables of
the theory, we must take into account the constraints (24) between the
variations. This can be done by means of Lagrange's method of undetermined
multipliers. If we denote the undetermined multipliers by
$\lambda^{\mu\nu}(\sigma,\tau)$, we find that the variation of $S$ takes the 
form:
$$
\eqalign{\delta S &= -{1\over{8\pi}}{1\over{3!}}\delta
A_{\alpha\beta\gamma}\phi^{\alpha\beta\gamma}\cr
 &+ \sum_{\sigma,\tau}
\lambda^{\mu\nu}(\sigma,\tau)
[det({\hbox{\bf M}}(\tau))C^{\alpha\beta\gamma}_{\mu\nu}(\sigma)-
det({\hbox{\bf M}}(\sigma))C^{\alpha\beta\gamma}_{\mu\nu}(\tau)]\delta
A_{\alpha\beta\gamma}.\cr}\eqno(26)
$$            
where we have used the well-known identity
$$
A_{\alpha\beta\gamma}\delta \phi^{\alpha\beta\gamma} = 0.\eqno(27)
$$
In Eq.(26), the sum is taken over all four-simplexes $\sigma$ and $\tau$ of the
complex. When the variation of $S$ has been written as in Eq.(26), the variations
$\delta A_{\alpha\beta\gamma}$ can be considered, formally, as independent
variations, and the condition $\delta S=0$ implies:
$$
\phi^{\alpha\beta\gamma} = 
48\pi\sum_{\sigma,\tau}\lambda^{\mu\nu}(\sigma,\tau)
[det({\hbox{\bf M}}(\tau))C^{\alpha\beta\gamma}_{\mu\nu}(\sigma)-
det({\hbox{\bf M}}(\sigma))C^{\alpha\beta\gamma}_{\mu\nu}(\tau)].\eqno(28)
$$            
At this point we use Eq.(18), and we get:
$$
M_{\alpha\beta\gamma}^{\mu\nu}\phi^{\alpha\beta\gamma} = 0,\eqno(29)
$$
which, according to Eq.(10) means that
$$
{{\partial A_{\alpha\beta\gamma}}\over{\partial
s_{\mu\nu}}}\phi^{\alpha\beta\gamma} = 0.\eqno(30)
$$
In other words, we have obtained the Einstein-Regge equations in vacuum. When
obtaining these equations, we did not vary the edge lengths but we varied the
areas, and we imposed the constraints (24) between their variations. It should
be noted that when obtaining the constraints (24) we assumed the determinants
of the matrices ${\hbox{\bf M}}(\sigma)$ and ${\hbox{\bf M}}(\tau)$ to be
non-zero. However, the constraints (24) are still valid even if one, or both,
of the determinants $det({\hbox{\bf M}}(\sigma))$ and $det({\hbox{\bf
M}}(\tau))$ are zero. In particular, the constraints (24) imply the Einstein-Regge
equations no matter whether the determinants $det({\hbox{\bf
M}}(\sigma))$ corresponding to the four-simplexes $\sigma$ of the complex are
zero or non-zero.

          In this paper we have investigated the constraints between the area
variables in Regge calculus. We did not find the constraints between the area
variables themselves, but we did find the constraints between their variations.
We showed that when the action of Regge calculus is varied with respect to the
area variables, and the constraints are taken into account, the Einstein-Regge
equations are recovered. Hence, it appears that it is possible to use areas of
two-simplexes, instead of lengths of edges, as the dynamical variables of Regge
calculus, provided that appropriate constraints are imposed.

          The key point in our derivation of the constraints was an observation
that the number of two-simplexes of a four-simplex equals with the number of
its edges. This property of four-simplexes enables one to express the
variations of the squared edge lengths of a four-simplex in terms of the
variations of the areas of its two-simplexes. This implies that when an edge is
shared by two four-simplexes, the variation of its squared length can be
written in two different ways in terms of the variations of the area variables.
The uniqueness of the variations of the squared edge lengths then gives the
constraints between the variations of the areas. However, the number of
constraints thus gained is enormous: for every edge shared by $n$
four-simplexes there are
$$
\left(\matrix{n\cr
              2\cr}\right)
$$ 
constraints, and hence all of the constraints (24) cannot be linearly
independent of each other. It remains to be seen, whether this deficiency in
our analysis can be removed by further study.

\bigskip

\centerline{\bf Acknowledgments}

\bigskip

I thank Markku Lehto for his constructive criticism during the preparation of
this paper and Pasi Repo for calculating the matrix equation (20).

\bigskip

\bigskip

\centerline{\title References}

\bigskip

[1] T. Regge, Nuovo Cimento, {\bf 19} (1961), 558

\medskip

[2] C. W. Misner, K. Thorne and J. A. Wheeler, {\it Gravitation} (Freeman, San

Fransisco, 1973)

\medskip

[3] R. M. Williams and P. Tuckey, Class. Quant. Grav. {\bf 9} (1992), 1409

\medskip

[4] For the most recent review, see  C. Rovelli, gr-qc/9710008

\medskip

[5] C. Rovelli, Phys. Rev. {\bf D48} (1993), 2702

\medskip

[6] J. M\"{a}kel\"{a}, Phys. Rev. {\bf D49} (1994), 2882

\medskip

[7] J. W. Barrett, M. Rocek and R. M. Williams, gr-qc/9710056

\bye